\documentclass[a4paper]{jpconf}
\usepackage[utf8]{inputenc} 
\usepackage{graphicx}
\usepackage{amsmath}
\usepackage{cite}

\newcommand{\beq}{\begin{equation}}
\newcommand{\eeq}{\end{equation}}
\newcommand{\beqa}{\begin{eqnarray}}
\newcommand{\eeqa}{\end{eqnarray}}
\DeclareMathOperator{\im}{Im}

\begin{document}
\title{Dilepton Production in Transport Calculations and Coarse-Grained Dynamics }
\author{Stephan Endres$^{1, 2}$, Hendrik van Hees$^{1, 2}$, Janus Weil$^{2}$ and Marcus Bleicher$^{1, 2}$}
\address{$^{1}$Institut f\"{u}r Theoretische Physik, Universit\"{a}t Frankfurt, Max-von-Laue-Stra{\ss}e 1, 60438 Frankfurt, Germany}
\address{$^{2}$Frankfurt Institute for Advanced Studies, Ruth-Moufang-Stra{\ss}e 1, 60438 Frankfurt, Germany}
\ead{endres@th.physik.uni-frankfurt.de}
\begin{abstract}
We present transport calculations with the Ultra-relativistic Quantum Molecular Dynamics approach (UrQMD) for dilepton spectra at SIS energies. While we obtain a good agreement with experiment for elementary reactions, in heavy-ion collisions an excess in the invariant mass spectra is observed which cannot be described by the model. As the pure transport calculations do not include any in-medium effects and are limited to hadronic degrees of freedom, we present an alternative approach that uses coarse-grained output from transport calculations to determine thermal dilepton emission rates. For this we apply the medium-modified $\rho$ spectral function by Eletsky et al. In a first exemplary comparison to data from the NA60 experiment we find that the coarse-graining approach gives reasonable results.
\end{abstract}
\section{Introduction}
Dileptons are a unique tool to study the properties of hot and dense matter created in nuclear collisions. They might serve as probes for the in-medium properties of vector mesons and the predicted restoration of chiral symmetry \cite{Xia:1988ym,Rapp:1999ej,Leupold:2009kz}. Unlike hadrons, the lepton pairs produced in the nuclear fireball do not participate in the strong interaction and therefore penetrate the strongly interacting medium with negligible final-state reactions. Thus we gain insight into all the different stages of a nuclear collision, from the first nucleon-nucleon interactions to the final freeze-out. But this also means that in experimental measurements we obtain time-integrated spectra only, stemming from a broad variety of sources. In consequence we need good models that help to understand the production mechanisms and their contribution to the total spectra.
\\
Especially at lower bombarding energies, where the dominant dilepton contribution originates from hadronic decays, transport models have been successful in describing the experimentally measured dilepton spectra \cite{Weil:2012ji, Bratkovskaya:2007jk, Schmidt:2008hm, Vogel:2007yu}. In these proceedings we present recent results from our calculations with the UrQMD model at SIS energies. However, for a hot and dense evironment as created in ultra-relativistic heavy-ion collisions it is supposed that medium effects play a crucial role for dilepton production, but it is highly difficult to implement them in a transport model. We argue here that a coarse-graining approach is a good way to apply in-medium spectal functions within an underlying microscopic description of the reaction dynamics. 
\section{Transport Approach}
\begin{figure}[t]
\begin{minipage}{19pc}
\includegraphics[width=20.0pc,height=15.1pc]{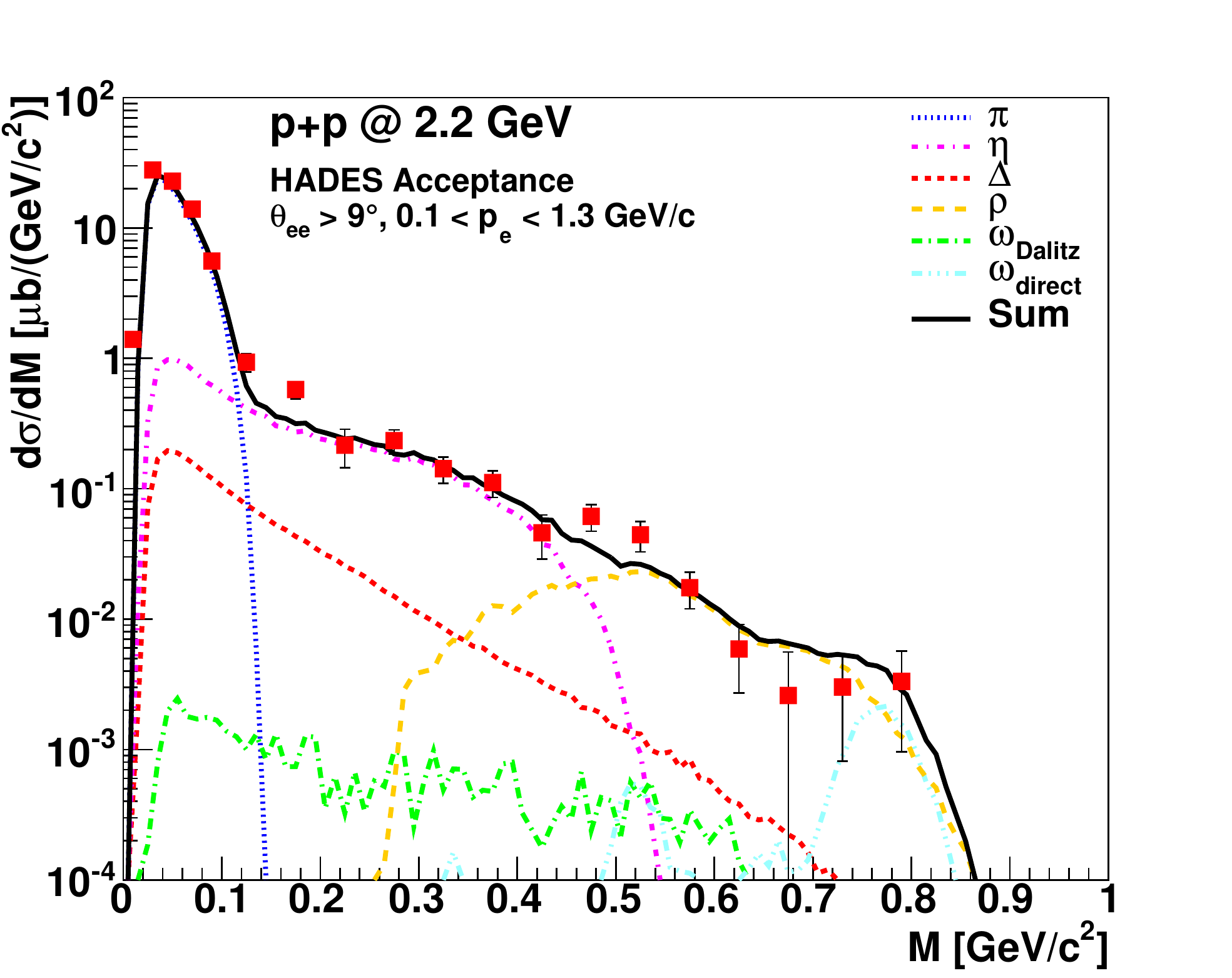}
\end{minipage}
\begin{minipage}{19pc}
\includegraphics[width=20.0pc,height=15.1pc]{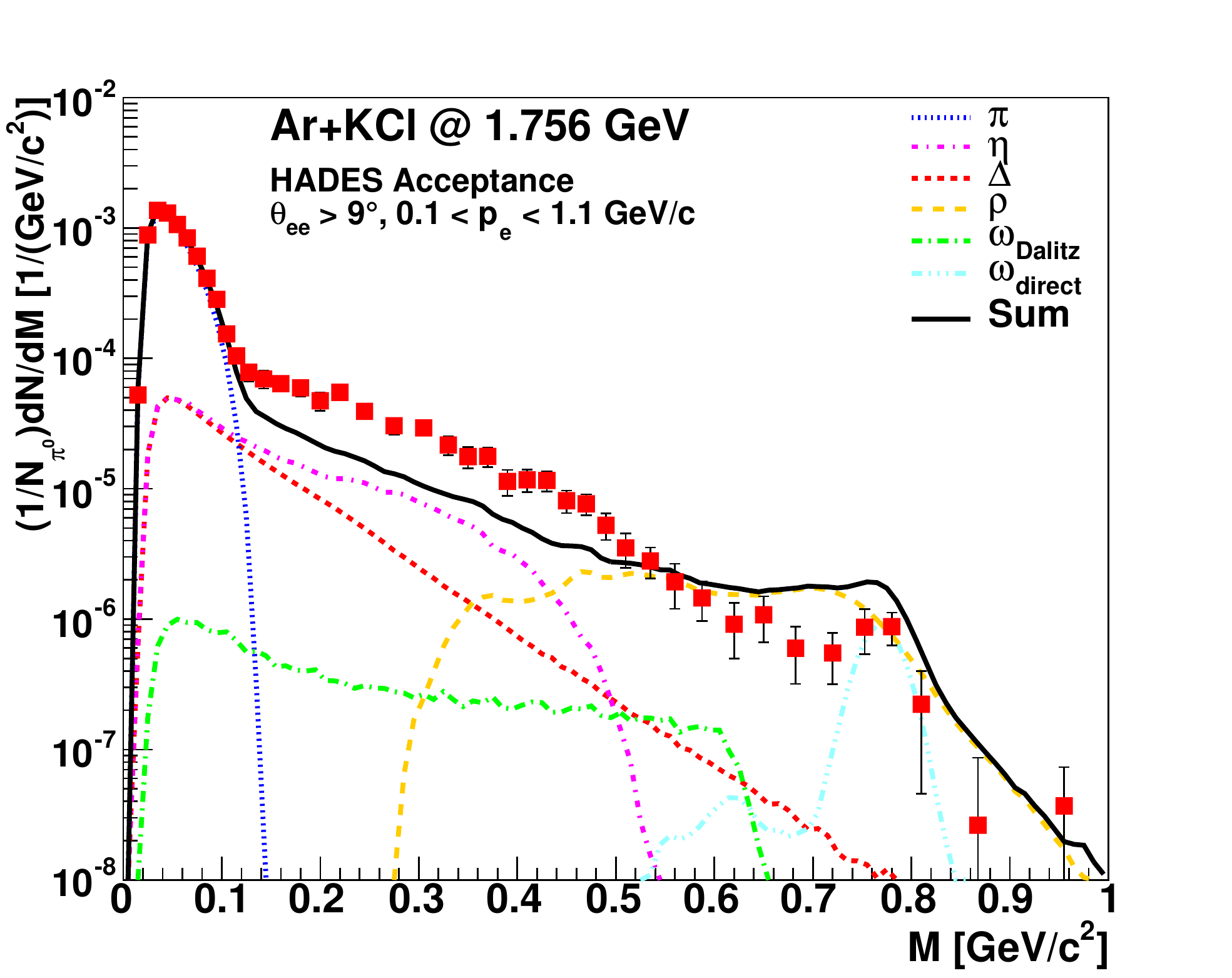}
\end{minipage}%
\caption{\label{fig1} Invariant mass spectra for p+p at 2.2 GeV (left)  and Ar+KCl at 1.756 AGeV (right) obtained with the UrQMD transport model. The results are corrected for the HADES acceptance
and compared with data from the HADES Collaboration \cite{Agakishiev:2011vf, Agakishiev:2012tc}.
}
\end{figure}
For our calculations of dilepton spectra we use the Ultra-relativistic Quantum Molecular Dynamics model (UrQMD), which is a hadronic non-equilibrium transport approach and includes all baryons and mesons with masses up to 2.2 GeV \cite{Bass:1998ca, Bleicher:1999xi, Petersen:2008kb}. In this model the production of resonances takes place either via inelastic collisions or the decay of higher resonances. As dilepton-production channels the Dalitz decays of the $\pi^{0},\ \eta$ and $\omega$ meson and of the $\Delta(1232)$ resonance are included, as well as the direct decay channels of the $\omega,\ \rho$ and $\phi$ vector mesons. The Dalitz decays are treated in a two-step process, first the decay into a virtual photon and as second step the electromagnetic conversion into a lepton pair. In contrast to previous calculations we now apply the newer parametrization for the radiative decay width of the $\Delta(1232)$ by Krivoruchenko \cite{Krivoruchenko:2001hs}. Dalitz decays of other baryonic states than the $\Delta(1232)$ are not treated explicitly, but these resonances contribute to the dilepton spectra via their decays into light mesons. For further details on dilepton calculations with UrQMD we refer the reader to \cite{Schmidt:2008hm, Endres:2013nfa}.
\\
There are several parameters that strongly influence the dilepton production within our model: Especially the number and types of baryonic resonances which are included, their production cross-sections and their branching ratios into light mesons. These parameters have to be adapted to experimental constraints. However, in case of the resonance production cross-sections the data are very limited, and almost no information is available in the HADES energy regime up to $\sqrt{s}=$\ 3\ GeV \cite{Flaminio:1984gr}. The situation is not better for many decay-branching ratios: When comparing the results of different partial-wave analyses one finds big differences as well \cite{Manley:1992yb, Anisovich:2011fc}.  We reevaluated all the available data and adjusted some of the branching ratios and $N^{*}$ and $\Delta^{*}$ production cross-sections within the model to get the best possible consistency with the available experimental data, e.g.\ the inclusive and exclusive resonance procuction cross-sections.
\\
The results presented in Figure 1 show the invariant mass spectra of dileptons produced in proton-proton reactions at 2.2 GeV (left plot) and in Ar+KCl at 1.756 AGeV (right plot) compared to HADES data \cite{Agakishiev:2011vf,Agakishiev:2012tc}. We see that the description of elementary reactions is in a good agreement with experiment, while in the larger system Ar+KCl we find on the one side that the $\rho$ contribution in our calculation overshoots the data in the pole mass region and on the other hand that the experimental results show an excess between 150 and 500 MeV mass which is not reproduced with our model. There may be several possible explanations for this excess, e.g.\ bremsstrahlung effects or medium modifications of the vector mesons. However, a different study with the GiBUU model \cite{Buss:2011mx}, which includes the bremsstrahlung contribution in soft-photon approximation, still underestimates the yield in this low-mass region \cite{Weil:2012yg}.
\section{Coarse-Graining Approach}
\begin{figure}[t]
\begin{minipage}{19pc}
\includegraphics[height=13.0pc]{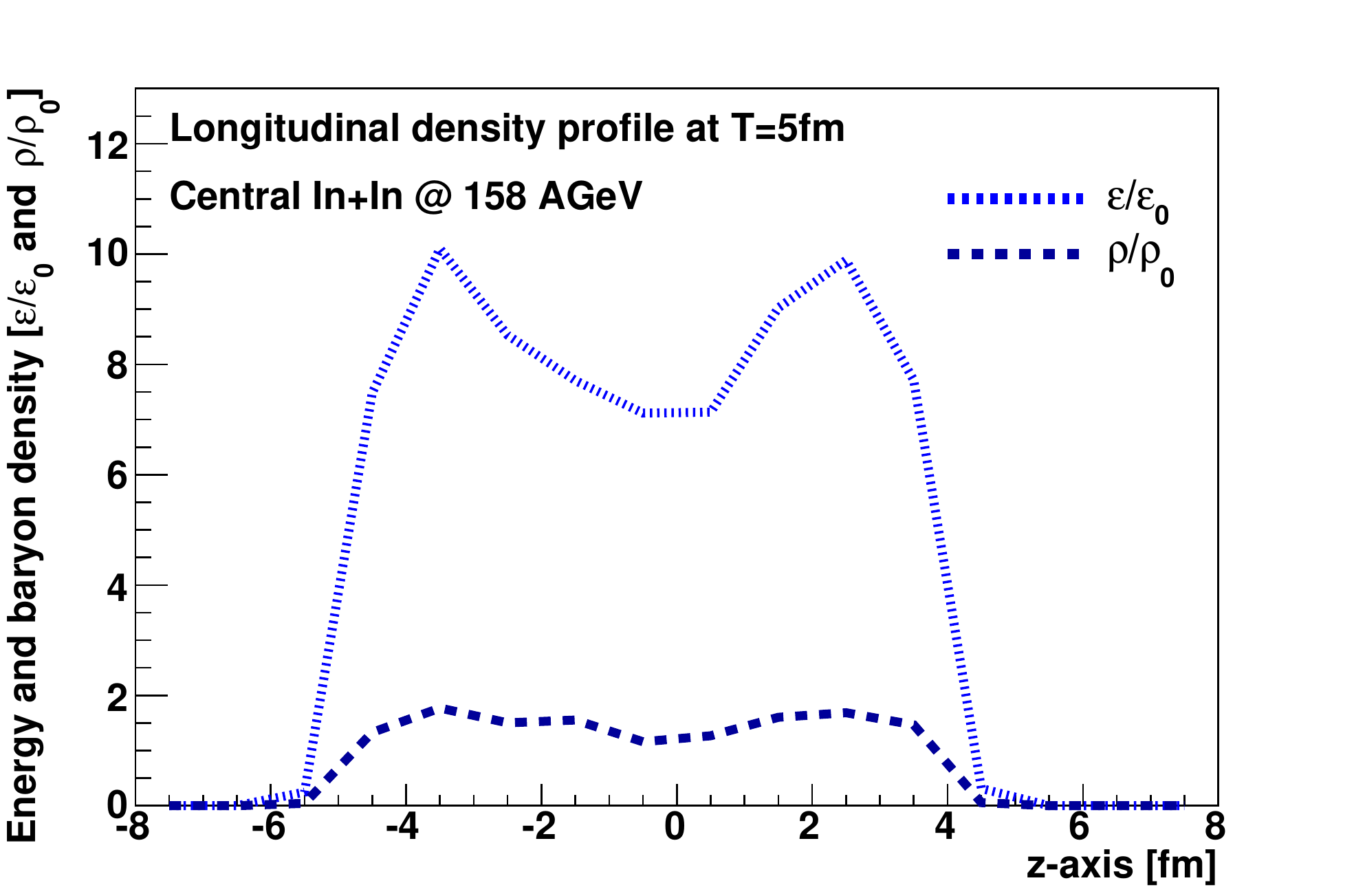}
\end{minipage}
\begin{minipage}{19pc}
\includegraphics[height=13.0pc]{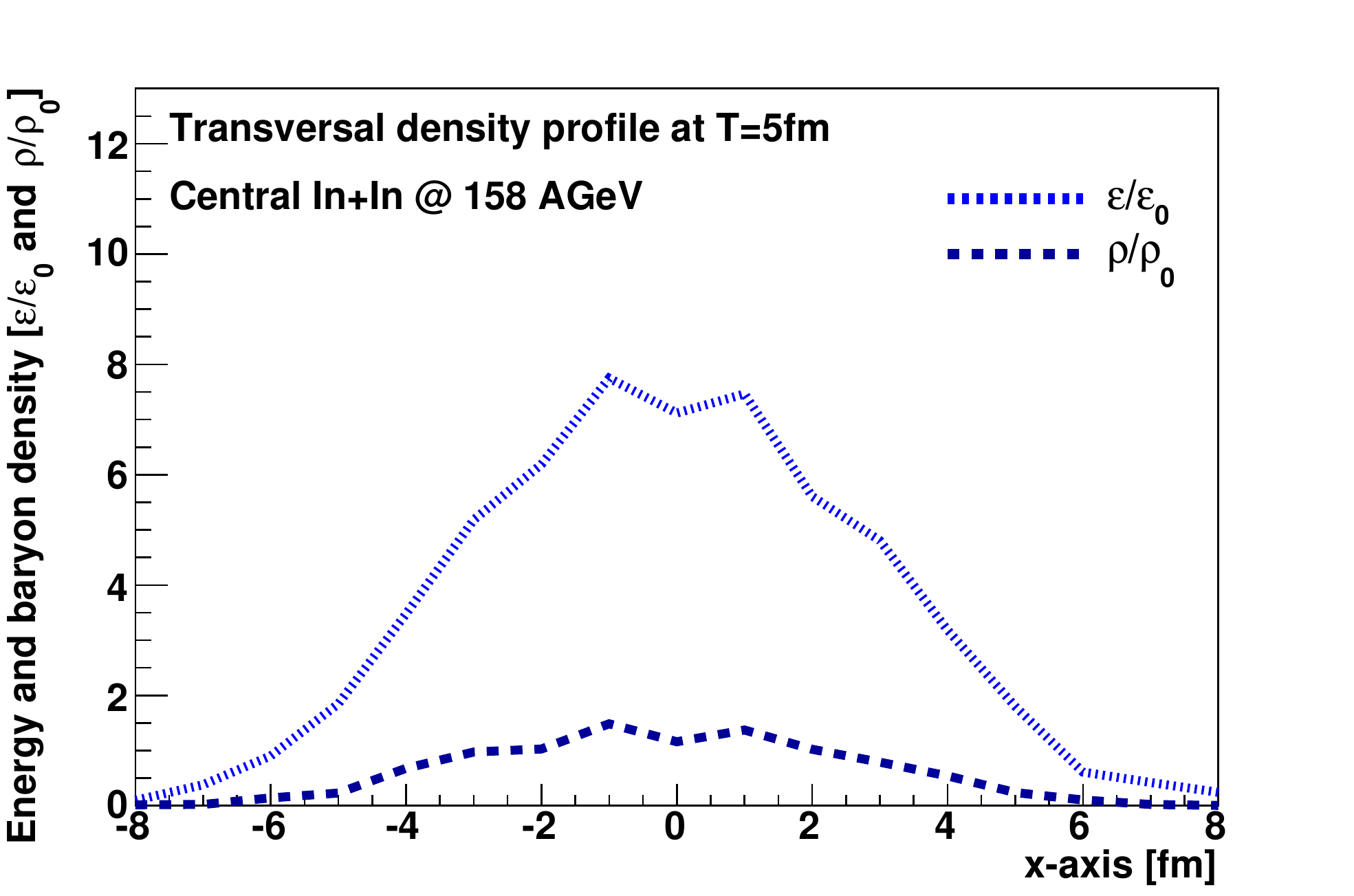}
\end{minipage}%
\caption{\label{fig2} (Left) Longitudinal profile (i.e.\ along the beam axis) for the energy and baryon desities in a central In+In collisions at 158 AGeV, as obtained from the coarse-graining of UrQMD events. The values here are given in units of the nuclear-matter ground state energy and baryon densities. (Right) The same plot but for the transverse energy and baryon densities.}
\end{figure}
It is important to note that our present transport calculations do not include any kind of explicit in-medium modifications (besides reabsorption and rescattering, which are dynamically implemented). For an entirely proper calculation it would be necessary to include off-shell effects and multi-particle interactions, which is a highly non-trivial challenge. Another problem arises at higher bombarding energies, where dilepton emission from thermal radiation comes into play, as the UrQMD model only includes hadronic degrees of freedom. One method to avoid these problems is the integration of dilepton radion from an intermediate hydrodynamic phase within a hybrid model, using in-medium dilepton rates \cite{Santini:2011zw}. In the following we sketch a different approach that uses coarse-grained transport dynamics instead of hydrodynamics to calculate in-medium emission rates. A similar procedure was first proposed in \cite{Huovinen:2002im}.
\\
As first step we put the UrQMD output on a 3+1 dimensional space-time grid and determine baryon and energy density for each cell as an average over several hundred events. Assuming thermal equilibrium we transform into the local rest frame of the cell, according to Eckart's definition that requires the net baryon flow to be zero. Then we apply an equation of state to extract temperature $T$ and baryon chemical potential $\mu_{B}$. For SPS and higher energies we use a chiral equation of state which includes chiral symmetry restoration and a deconfinement phase transition. It shows a crossover between the hadronic and QGP phase with a coexistence region \cite{Steinheimer:2010ib}.
\\
Assuming vector meson dominance we can calculate the equlibrium $l^{+}l^{-}$ emission rates per volume and momentum in each cell according to \cite{Rapp:1999ej} as 
\begin{equation}
\frac{\mathrm{d}^8 N_{V\rightarrow ll}}{\mathrm{d}^4 x \mathrm{d}^4 q}=-\frac{\alpha^2m_V^4}{\pi^3 g_V^2}\frac{L(M^2)}{M^2}f_B(q_0;T)
\im D_V^{(\mathrm{ret})}(M,q;T,\mu_B), \
\end{equation}
in terms of the imaginary part of the meson's retarded propagator $\im D_V^{(\mathrm{ret})}$. Here $L(M^{2})$ accounts for the lepton phase space and the function $f_{B}$ is the Bose-Einstein distribution. As the largest in-medium contribution is expected to stem from the $\rho$ meson we concentrate on this contribution in our considerations. For a complete study it will be necessary to include the in-medium modifications of the $\omega$ and the $\phi$ meson, too.
\\
Regarding the $\rho$ in-medium spectral function several approaches exist, relying either on calculations from hadronic many-body theory or on empirical scattering amplitudes obtained from experiment. Here we apply the description by Eletsky and Kapusta \cite{Eletsky:2001bb}, which uses experimental data in linear density approximation to include the scattering from pions and nucleons in the vector meson's retarded self-energy, i.e. $\Sigma_{\rho}=\Sigma^{\mathrm{vac}}+\Sigma^{\rho N}+\Sigma^{\rho \pi}$. These additional effects on the self energy result in a broadening of the spectral width in a hot and dense medium.
\\
At high collision energies, 
\begin{figure}[t]
\begin{minipage}{19pc}
\includegraphics[height=13.0pc]{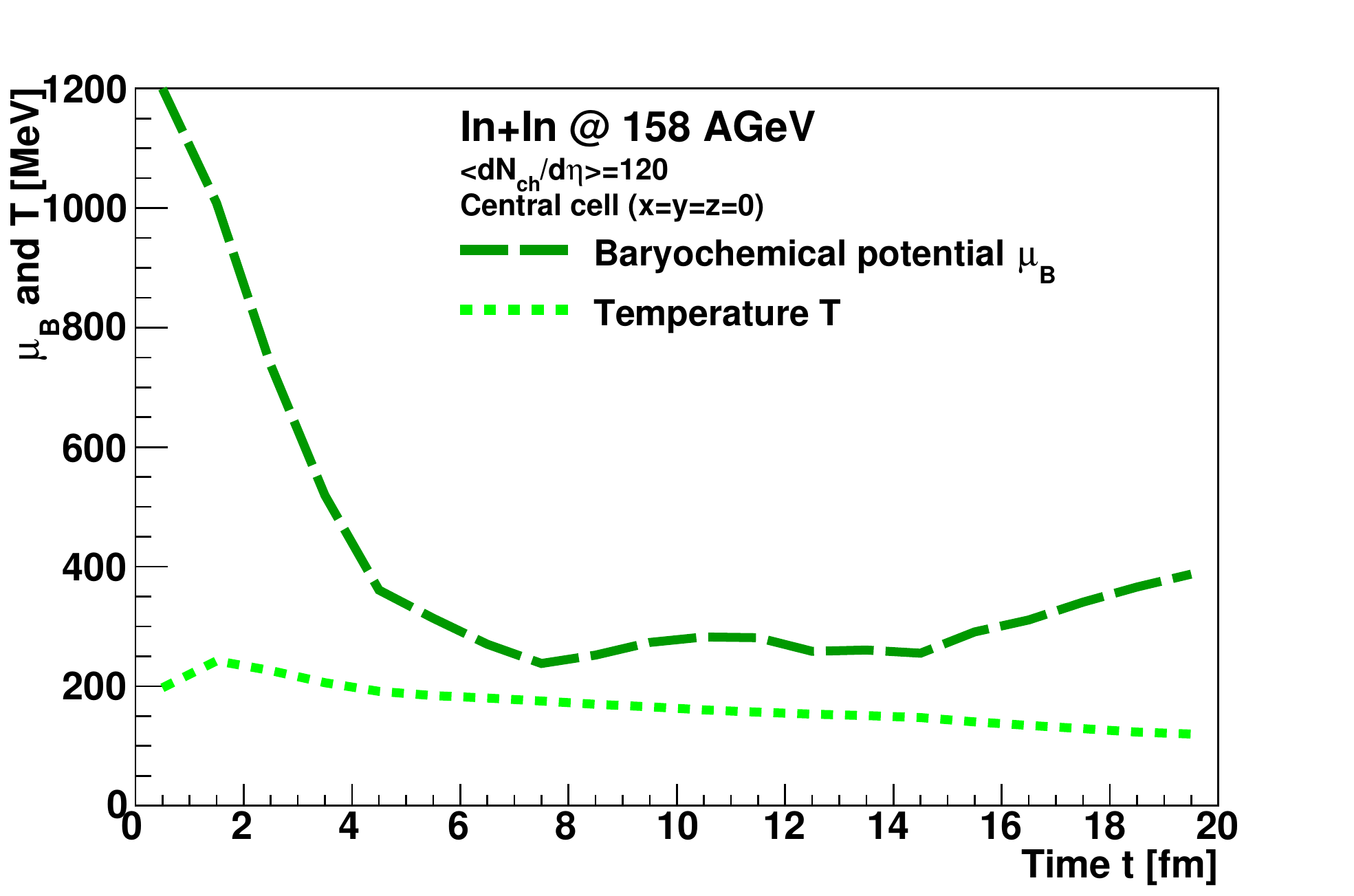}
\end{minipage}
\begin{minipage}{19pc}
\includegraphics[height=13.0pc]{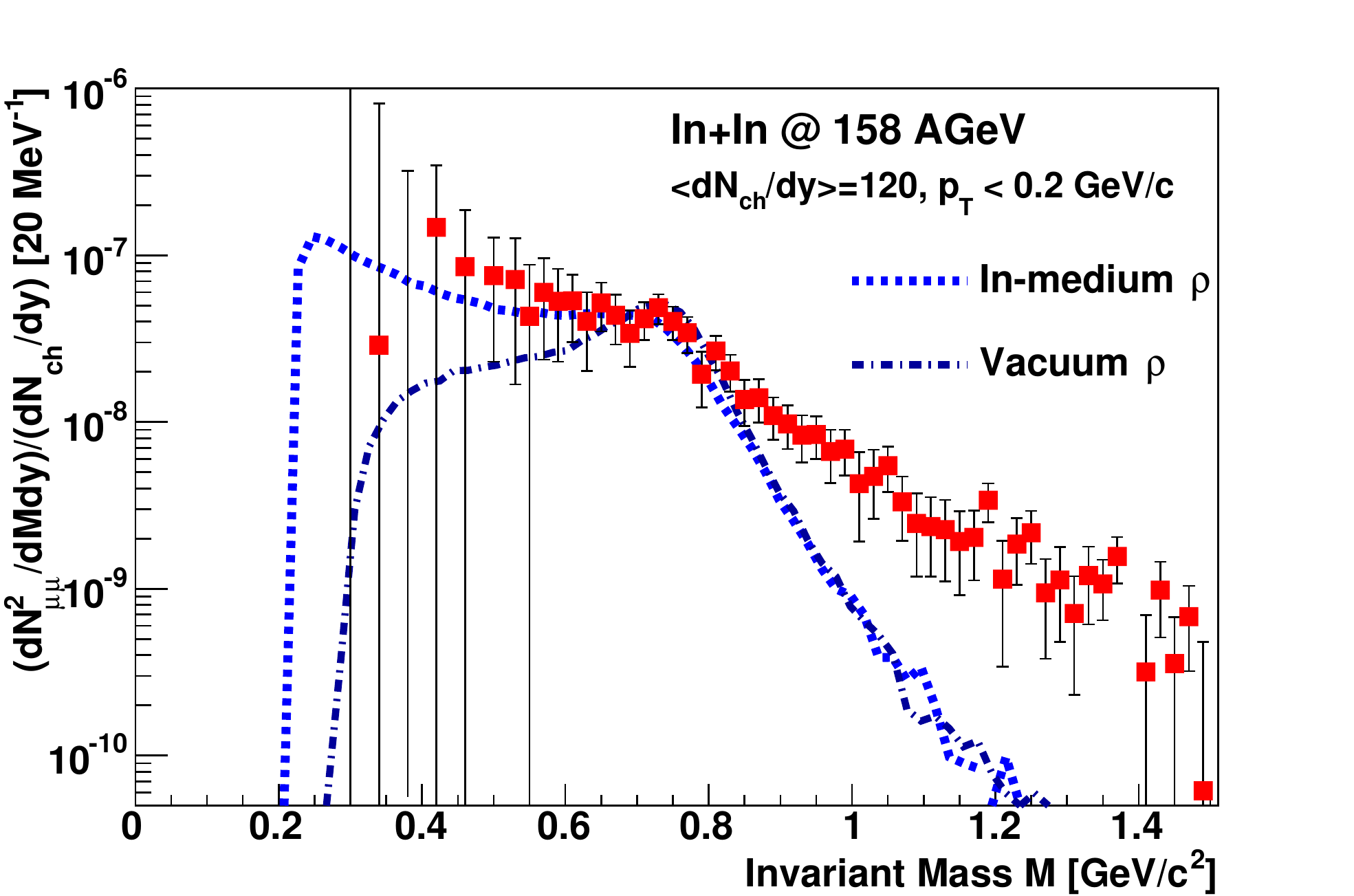}
\end{minipage}%
\caption{\label{fig3} (Left) Time evolution of the baryochemical potential $\mu_{B}$ and temperature $T$ for a central cell in our grid of In+In collisons at 158 AGeV as obtained with a chiral equation of state. (Right) Dimuon excess spectrum from coarse-grained dynamics for In+In at $E_{lab}$ = 158 AGeV with d$N_{ch}$/d$y$ = 120, here in the $p_{t}<0.2$ GeV region. The in-medium $\rho$ and, for comparison, the pure vacuum contribution are shown. The results are compared to NA60 data \cite{Arnaldi:2008fw}.
}
\end{figure}
is is not sufficient to look only at the modifications of the vector mesons, but also additional thermal sources must be taken into account. If a deconfined phase is created during the heavy-ion collision we will also find a contribution from $q\bar{q}$ annihilation in the Quark-Gluon Plasma. The rates can be calculated using the electromagnetic cross-section for the annihilation of a quark-antiquark pair into dileptons \cite{Cleymans:1986na}. Similarly we expect multi-particle interactions in the medium. While two-pion interactions are included in the $\rho$ spectral function, it has been shown that also 
four-pion reactions play a crucial role for the production of lepton pairs \cite{Huang:1995dd}. However, as these contributions will become dominant only above 1 GeV in the invariant mass spectra we postpone a detailed study to a future work.
\\
The advantage of the coarse-graining approach described above is that we can simulate a realistic evolution of temperature and baryochemical potential during the nuclear collision related to the underlying microscopic dynamics. This is an important aspect as only a reliable description of the different stages enables to understand the role of the different dilepton-production channels. 
\\
To check our approach we make a first comparison with NA60 data on dimuon production in In+In collisions at 158 AGeV \cite{Arnaldi:2008fw}. In Figure 2 the longitudinal and the transverse profile of energy and baryon density ($\epsilon$ respectively $\rho_{B}$) are shown, resulting from the coarse graining of central UrQMD events at $t=$\ 5\ fm after the beginning of the collision. At this time the nuclei have already passed through each other as indicated by the two peak structures in the profile along the beam axis. In the center of the collision we find a hot and dense region with values of $\epsilon$ and $\rho_{B}$ reaching up to several times the ground-state densities. The resulting time evolution of temperature and baryon chemical potential is shown in Figure 3 (left), here exemplarily for a cell at the center of the collision. The baryon chemical potential decreases very fast after the beginning of the collision and then remains at a level of 300-400 MeV, while the temperature shows an initial peak around 250 MeV followed by a cooling down to around 100 MeV at $t=$\ 20\ fm. These results are in line with similar previous studies \cite{Bass:1998qm, Huovinen:2002im}.
\\
Looking at the resulting dilepton spectrum (here for $p_{t}<0.2$ GeV) in the right plot of Figure 3, we see that the low-mass tail of the excess spectrum (with all hadronic cocktail contributions already subtracted) is nicely filled up by a broadened in-medium $\rho$. Applying only a vacuum spectral function, in comparison, underestimates the contribution in the region below the $\rho$ pole mass significantly. Above a mass of 1 GeV the $\rho$ no longer fills up the spectrum, here one expects other thermal sources to be dominant (mainly QGP and four-pion emission) that are not included in the present calculations.
\section{Summary \& Outlook}
In summary we showed that our UrQMD transport calculations give a good description of experimental dilepton spectra, but the lack of a correct implementation of medium modifications remains a problem. To account for this we proposed that a coarse-graining approach might serve to obtain a realistic description of the reaction dynamics and the resulting dilepton rates. Our exemplary comparison to the data from the NA60 experiment shows a good agreement. Further studies have to be performed, e.g.\ by investigating different equations of state and spectral functions. Of special interest would be a check of the many-body calculation for the $\rho$ spectral function by Rapp and Wambach \cite{Rapp:1997fs}, which has proven successful in describing the dilepton results from NA60 and RHIC in a fireball model \cite{vanHees:2007th}, within our coarse-graining approach. 
\ack
We thank the HADES Collaboration, especially T. Galatyuk and C. Sturm, for providing their acceptance filters and data. This work was supported by BMBF, HIC for FAIR and H-QM.

\section*{References}
\begin{thebibliography}{28}
\bibitem{Xia:1988ym} 
Xia~L~H, Ko~C~M, Xiong~L and Wu~J~Q 1988
\textit{Nucl.\ Phys.} A {\bf 485} 721
\bibitem{Rapp:1999ej} 
Rapp~R and Wambach~J 2000
\textit{Adv.\ Nucl.\ Phys.} {\bf 25} 1
(\textit{Preprint} hep-ph/9909229)
\bibitem{Leupold:2009kz}
Leupold S, Metag V and Mosel U 2010
\textit{Int.\ J.\ Mod.\ Phys.}\ E {\bf 19} 147
(\textit{Preprint} arXiv:0907.2388 [nucl-th])
\bibitem{Weil:2012ji} 
Weil~J {\it et al.} 2012
\textit{Eur.\ Phys.\ J.} A {\bf 48} 111; Erratum-ibid. 2012 A {\bf 48} 150
(\textit{Preprint} arXiv:1203.3557 [nucl-th])
\bibitem{Bratkovskaya:2007jk} 
Bratkovskaya E L and Cassing W 2008
\textit{Nucl.\ Phys.}\ A {\bf 807}, 214 
(\textit{Preprint} arXiv:0712.0635 [nucl-th])
\bibitem{Schmidt:2008hm} 
Schmidt~K, Santini~E {\it et al.} 2009
\textit{Phys.\ Rev.} C {\bf 79} 064908
(\textit{Preprint} arXiv:0811.4073 [nucl-th])
\bibitem{Vogel:2007yu} 
Vogel~S, Petersen~H, \textit{et al.} 2008
\textit{Phys.\ Rev.} C {\bf 78} 044909
(\textit{Preprint} arXiv:0710.4463 [hep-ph])
\bibitem{Bass:1998ca}
Bass~S~A {\it et al.} 1998
\textit{Prog.\ Part.\ Nucl.\ Phys.} {\bf 41} 255
(\textit{Preprint} nucl-th/9803035)
\bibitem{Bleicher:1999xi}
Bleicher~M {\it et al.} 1999
\textit{J.\ Phys.} G {\bf 25} 1859
(\textit{Preprint} hep-ph/9909407)
\bibitem{Petersen:2008kb} 
Petersen~H, Bleicher~M, Bass~S~A and Stoecker~H 2008
\textit{Preprint} arXiv:0805.0567 [hep-ph]
\bibitem{Krivoruchenko:2001hs}
Krivoruchenko M I and Faessler A 2002
\textit{Phys.\ Rev.}\ D {\bf 65} 017502
(\textit{Preprint} nucl-th/0104045)
\bibitem{Endres:2013nfa}
Endres S and Bleicher M 2013
\textit{J.\ Phys.\ Conf.\ Ser.}\ {\bf 426} 012033
\bibitem{Flaminio:1984gr} 
Flaminio~V, Moorhead~W~G, Morrison~D~R~O and Rivoire~N 1984
\textit{CERN-HERA-84-01}
\bibitem{Manley:1992yb}
Manley D~M~ and  Saleski~E~M 1992
\textit{Phys.\ Rev.}\ D {\bf 45} 4002
\bibitem{Anisovich:2011fc}
Anisovich~A~V {\it et al.} 2012
\textit{Eur.\ Phys.\ J.}\ A {\bf 48} 15
(\textit{Preprint} arXiv:1112.4937 [hep-ph])
\bibitem{Agakishiev:2011vf} 
Agakishiev~G {\it et al.}  (HADES Collab.) 2011
\textit{Phys.\ Rev.} C {\bf 84} 014902
(\textit{Preprint} arXiv:1103.0876 [nucl-ex])
\bibitem{Agakishiev:2012tc} 
Agakishiev~G {\it et al.}  (HADES Collab.) 2012
\textit{Phys.\ Rev.} C {\bf 85} 054005
(\textit{Preprint} arXiv:1203.2549 [nucl-ex])
\bibitem{Buss:2011mx}
Buss O, Gaitanos T, Gallmeister K {\it et al.} 2012
\textit{Phys.\ Rept.}\ {\bf 512} 1
(\textit{Preprint} arXiv:1106.1344 [hep-ph])
\bibitem{Weil:2012yg}
Weil J and Mosel U 2013
\textit{J.\ Phys.\ Conf.\ Ser.}\ {\bf 426} 012035
(\textit{Preprint} arXiv:1211.3761 [nucl-th])
\bibitem{Santini:2011zw} 
Santini~E, Steinheimer~J \textit{et al.} 2011
\textit{Phys.\ Rev.} C {\bf 84} 014901
(\textit{Preprint} arXiv:1102.4574 [nucl-th])
\bibitem{Huovinen:2002im}
Huovinen P, Belkacem P {\it et al.} 2002
\textit{Phys.\ Rev.}\ C {\bf 66} 014903
(\textit{Preprint} nucl-th/0203023)
\bibitem{Steinheimer:2010ib}
Steinheimer~J, Schramm~S and Stocker H 2011
\textit{J.\ Phys.}\ G {\bf 38} 035001
(\textit{Preprint} arXiv:1009.5239 [hep-ph])
\bibitem{Eletsky:2001bb}
Eletsky V L {\it et al.} 2001
\textit{Phys.\ Rev.}\ C {\bf 64} 035202
(\textit{Preprint} nucl-th/0104029)
\bibitem{Cleymans:1986na}
Cleymans J, Fingberg J and Redlich K 1987
\textit{Phys.\ Rev.}\ D {\bf 35} 2153
\bibitem{Huang:1995dd}
Huang Z 1995
\textit{Phys.\ Lett.}\ B {\bf 361} 131
(\textit{Preprint} hep-ph/9506399).
\bibitem{Arnaldi:2008fw}
Arnaldi R {\it et al.}  (NA60 Collab.) 2009
\textit{Eur.\ Phys.\ J.}\ C {\bf 61} 711
(\textit{Preprint} arXiv:0812.3053 [nucl-ex])
\bibitem{Bass:1998qm}
Bass S A, Weber H, Ernst C, Bleicher M {\it et al.} 1999
\textit{Prog.\ Part.\ Nucl.\ Phys.}\ {\bf 42} 313
(\textit{Preprint} nucl-th/9810077)
\bibitem{Rapp:1997fs} 
Rapp~R, Chanfray~G and Wambach~J 1997
\textit{Nucl.\ Phys.}\ A {\bf 617} 472
(\textit{Preprint} hep-ph/9702210)
\bibitem{vanHees:2007th}
van Hees H and Rapp R 2008
\textit{Nucl.\ Phys.}\ A {\bf 806} 339
(\textit{Preprint} arXiv:0711.3444 [hep-ph])
\end{thebibliography}
\end{document}